# Rapid Seismic Waveform Modeling and Inversion with Neural Operators

Yan Yang, Angela F. Gao, Kamyar Azizzadenesheli, Robert W. Clayton, and Zachary E. Ross

*Abstract*— Seismic waveform modeling is a powerful tool for determining earth structure models and unraveling earthquake rupture processes, but it is usually computationally expensive. We introduce a scheme to vastly accelerate these calculations with a recently developed machine learning paradigm called the neural operator. Once trained, these models can simulate a full wavefield at negligible cost. We use a U-shaped neural operator to learn a general solution operator to the 2D elastic wave equation from an ensemble of numerical simulations performed with random velocity models and source locations. We show that full waveform modeling with neural operators is nearly two orders of magnitude faster than conventional numerical methods, and more importantly, the trained model enables accurate simulation for velocity models, source locations, and mesh discretization distinctly different from the training dataset. The method also enables convenient full-waveform inversion with automatic differentiation.

## I. INTRODUCTION

THE seismic wave equation relates displacement fields to external forces and the density and elastic structure in the Earth. Solutions to the wave equation form the basis of ground shaking simulations of large earthquakes [1]–[3] and full waveform inversion for Earth's structure [4]–[6]. Due to the highly heterogeneous nature of the Earth, as exemplified by subduction zones and sedimentary basins, there are no exact analytical solutions for these wavefields. Instead, approximate solutions are made possible by approximating derivatives through discretized spatial and time or frequency domains. Finite difference methods (FDM) have been popular since the early 80s due to their relatively straightforward formulation [7]–[9]. The spectral-element method (SEM), a particular case of finite element methods (FEM), which was introduced to seismology in early 2000s, combined the flexibility of FEMs with the accuracy of spectral approaches [10]–[13]. These numerical solvers impose a tradeoff between resolution and computation speed, with the computational cost proportional to the fourth power of frequency [14]. Thus, the cost of wave simulation is a major barrier to using full-waveform techniques for seismic inversion and updating models of the subsurface with new data.

A number of machine learning-based methods have been proposed in the past few years to provide a faster alternative for tackling seismological problems such as signal denoising [15]–[17], event detection [18]–[20], and phase association [21], [22]. Deep neural networks have also recently been used to solve partial differential equations (PDEs), such as the Eikonal equation and wave equation [23]–[27]. These approaches to solving PDEs offer not only speedup in computational capabilities, but also low-memory overhead, differentiability, and on-demand solutions. Such advantages facilitate deep learning being used for seismic inversion [28]–[32]. However, one major limitation of these approaches is that the solutions generated by these models are dependent on the specific spatial and temporal discretization in the numerical simulation training set.

Recently, a paradigm named 'neural operator' was developed to address the mesh-dependent shortcoming of classical neural networks by creating a single deep learning model that can be applied to different discretizations [33]–[36]. This is made possible because neural operators can provably learn mappings between infinite-dimensional function spaces [37] and therefore are suitable for learning general solution operators to PDEs, which are valid even when the PDE coefficients (e.g. elastic properties) are varied. Since first introduced [33], a variety of neural operator models have been developed. In particular, the Fourier neural operator (FNO) is a model that uses the fast Fourier transform as an integral operator, and has been shown to outperform other neural operators in terms of efficiency and accuracy [38]. The FNO has been applied to many types of scientific problems including weather forecasting [39], $CO_2$ sequestration [40], and coastal flooding [41].

Within the domain of seismology, neural operators were also recently used to learn general solution operators to the 2-D acoustic wave equation, a simplified case of the elastic wave equation [42]. This pilot study demonstrated that it was possible for a single FNO model to predict a complete wavefield given an arbitrary velocity model and mesh discretization. The success of this limited case highlights the potential of these methods, however, extending the method from the acoustic wave equation to the elastodynamic case requires substantially increased model complexity. By comparison with neural networks, FNO is not considered to be a deep architecture, and is most analogous to the fully-connected neural networks employed heavily until the 2010s. A U-shaped neural operator (U-NO) was recently proposed to enable very deep neural operators and facilitate fast training, data efficiency, and hyperparameter selection robustness [43].

In this paper, we apply the U-NO architecture to full seismic waveform modeling. We train a U-NO model to learn a general solution operator to the 2D elastic wave equation and demonstrate that the trained model enables fast and accurate simulation for source locations, velocity structures and mesh discretization beyond the training dataset. The trained U-NO also allows for efficient full-waveform inversion with automatic differentiation.

## II. METHODS

### A. Neural operator learning

Operators are maps between function spaces, and the purpose of operator learning is to learn the operator given a dataset of input-output pairs. In seismology, it is common to write solutions to the wave equation, $U(x)$ in terms of a linear integral operator acting on a source function, $A(x)$,

$$U(x) = \int G(x,y) A(y) dy. \quad (1)$$

where $x \in \mathbb{R}^4$ is the physical domain and $G$ is a so-called Green's function defined for a particular velocity model. Equation 1 holds so long as the velocity model is not varied because the wave equation remains a linear operator.

Instead, if we consider the case where the input function, $A(x)$, is a velocity model, the solution operator, $\mathcal{L}$, relating this to $U(x)$ is nonlinear and cannot be written in the form of (1),

$$U(x) = (\mathcal{L} A)(x).$$

The most general version of the nonlinear solution operator $\mathcal{L}$ for the elastic wave equation is not known in closed form.

Neural operators are a class of models that aim to solve this problem, as they provably can learn a wide array of nonlinear operators. Their basic form consists of a composition of linear operators with nonlinear activations. More specifically, a neural operator with $L$ layers can be written as:

$$v_0(x) = (P A)(x),$$
$$v_{l+1}(x) = \sigma\big(W_l v_l(x) + \int \kappa_l(x,y) v_l(y) dy\big), l = 0, .., L-1$$
$$U(x) = (Q v_L)(x). \quad (2)$$

where $v_l$ is the input function at the $l^{\text{th}}$ layer, $P$ is a pointwise operator that lifts the input function to a higher dimensionality, $Q$ is a pointwise operator that projects the function back to the desired output dimensionality, $W_l$ is a linear pointwise transformation that can keep track of non-periodic boundary behavior, $\sigma$ is a pointwise nonlinear activation operator, and $\kappa_l$ is a kernel function that acts along with the integral as a global linear operator.

A neural operator is parameterized by $P$, $Q$, $W_l$, and $\kappa_l$. A critical aspect of this class of models is that these parameters are independent of the numerical discretization of the physical domain, i.e. they are shared across all possible discretizations in a similar way that in convolutional networks, the parameters are shared across neurons. It is this property that allows for the learning of maps between infinite dimensional function spaces, as the discretization can be chosen dynamically at inference time independently of what was used for training.

If we are given a dataset of $N$ numerical simulations, $\{A_i, U_i\}_{i=1}^N$, where the $A_i$ are chosen to span the range of the expected function space, we can train a neural operator in a supervised fashion to map from arbitrary $A$ into $U$.

Due to the expense of evaluating integral operators, neural operators may lack the efficiency of convolutional or recurrent neural networks in finite-dimensional settings. The FNO was proposed to mitigate this difficulty through the fast Fourier transform [38]. The kernel integral operator in (2) can be considered a convolution operator, defined in Fourier space as:

$$\int \kappa_l(x,y) v_l(y) dy = \mathcal{F}^{-1}(\mathcal{F}(\kappa_l) \cdot \mathcal{F}(v_l)). \quad (3)$$

where $\mathcal{F}$ and $\mathcal{F}^{-1}$ denote Fourier Transform and its inverse, respectively. However, FNO imposes that each layer is a map between functions spaces with identical domain spaces, which may cause a large memory usage. The U-NO, an analogy to the U-net architectures, was proposed to allow progressively transforming the input function space with respect to a sequence of varying domains [43], [44]. After the lifting operator $P$, a sequence of $L_1$ non-linear integral operators $G_i$ is applied to $v_0$ and map the input to a set of functions with decreasing dimensional domain. Then a sequence of $L_2$ non-linear integral operators $G_i$ is applied to $v_{L_1+1}$ and map the input to a set of functions with increasing dimensional domain before the projection operator $Q$. Skip connections [44] are included to add vector-wise concatenation of $v_{L_1+i}$ and $v_{L_1-i}$. The contracting and expanding parts are symmetric. The architecture of the U-NO used in this study is illustrated in Figure 1, and we refer the interested readers to the references [33]–[36], [38], [43] for more details.

### B. Numerical simulation

We set up a training dataset of random source locations, S-wave velocity ($V_S$) models, and P- to S-wave velocity ratios ($V_P/V_S$). We define the velocity model on a 64 × 64 mesh with 0.16 km grid spacing. The source is set as an isotropic explosive source randomly distributed on the mesh. The $V_S$ has an average background of 3 km/s and perturbed by random fields with a von Kármán covariance function with the following parameters: Hurst exponent $\kappa$=0.5, correlation length $a_x$=$a_y$=8 grids, and the fractional magnitude of the fluctuation $\varepsilon$=10% background velocity. The power spectral density function of the von Kármán type random field follows a power law (fractal randomness) and can accurately represent the distribution of Earth's heterogeneity [45]. The $V_P/V_S$ is simplified to an average background of 1.732 perturbed by a smooth Gaussian random field with the following parameters: correlation length $\lambda$=32 grids, standard deviation $\sigma$=2% background. This work, as our very first experiment to evaluate the feasibility of solving 2D elastic wave equations, wants to focus on the parameters that the wavefield is most sensitive to. Therefore, we use the empirical relation between density and $V_S$ to compute the density [46]. Other input parameters such as density and attenuation may be explored in future work. A total of 20,000 random sets of models are generated and each of them is input to a GPU based 2D finite difference code in Cartesian coordinates to simulate the 2-D displacement field [47]. For simulation, the top boundary is set with a free-surface boundary condition and the other three edges have absorbing boundary conditions. A total of 4-sec wavefield with a time step of 0.01 sec and a major frequency content up to 6 Hz is simulated. Each simulation takes about 1.23 sec with a GPU memory usage of 0.3 GB.

*C. U-NO model training*

We developed a framework that applies U-NO to the 2D elastic wave equation. The architecture is depicted schematically in Figure 1. U-NO takes the source location and $V_P$ and $V_S$ as inputs, where $V_P$ is calculated from $V_S$ and $V_P/V_S$. $V_P$ and $V_S$ are then passed through a point-wise lifting operator. A sequence of non-linear integral operators (encoders) are applied that gradually contract the physical domain size after each inverse Fourier transform step, while simultaneously increasing the number of channels in the co-domain. These operators are followed by a sequence of non-linear integral operators (decoders) that progressively expand the physical domain, and decrease the number of channels. Finally, a point-wise projection operator leads to the output function [43]. The output of the U-NO model is the complete horizontal and vertical displacement wavefield function over the medium domain, which can be queried at any mesh points desired, regardless of the input and output training mesh used.

We describe the detailed parameters used in U-NO below following the notations in Figure 1. The goal is to learn an operator mapping from the input function $a$ to the output function $u$. The training is on an input mesh of $X_{in} \times Y_{in} \times T_{in}$ and an output mesh of $X_{out} \times Y_{out} \times T_{out} \times C_{out}$, where $X_{in} = Y_{in} = X_{out} = Y_{out} = 64$, $T_{in} = 3$ representing source, $V_S$, and $V_P/V_S$ distribution on the mesh, $T_{out} = 128$ for 32 Hz data output, and $C_{out} = 2$ representing two displacement components (horizontal and vertical). This work applies the U-NO architecture designed for mapping between 3-D spatio-temporal function domains $(x, y, t)$ without any recurrent composition in time [43]. The fourth dimension $C_{out}$ of the output function $u$ can be created in the last step through the projection operator $Q$. Constructing the operator to learn the mapping between 3-D spatio-temporal function domains:

$G: \{a: [0,1]^2 \times [0, T_{in}] \to \mathbb{R}^{d_A}\}$
$\to \{u: [0,1]^2 \times [0, T_{out}] \to \mathbb{R}^{d_U}\}.$ (4)

The operators $\{G_i\}_{i=0}^L$ as shown in Figure 1 that are used to construct the U-NO are defined as:
$G_i: \{v_i: [0, \alpha_i]^2 \times \mathcal{T}_i \to \mathbb{R}^{d_{v_i}}\}$
$\to \{v_{i+1}: [0, c_i^s \alpha_i]^2 \times c_i^t \mathcal{T}_i \to \mathbb{R}^{c_i^c d_{v_i}}\}.$ (5)

where $[0, \alpha_i]^2 \times \mathcal{T}_i$ is the domain of the input function $v_i$ to the operator $G_i$, and $c_i^s, c_i^t$, and $c_i^c$ are the expansion or contraction factors for the spatial domain, temporal domain, and co-domain for $i^{th}$ operator, respectively. Note that $\mathcal{T}_0 = [0, T_{in}]$, $\alpha_0 = \alpha_{L+1} = 1$, and $\mathcal{T}_{L+1} = [0, T_{out}]$. In this work we set the number of layers to $L$=8. The details of the expansion and contraction factors $c_i^s, c_i^t$, and $c_i^c$ are in Figure 1. The lifting operator $P$ to convert the input to a higher dimension channel space is a fully-connected neural network with channel number $d_0 = 16$. The projection operator $Q$ to the output domain is also a fully connected neural network. The activation function used in each FNO block is the Gaussian Error Linear Unit (GELU) [48].

With the simulation dataset and the U-NO design, we train the U-NO model in a supervised manner with the objective of learning the general solution operator to the wave equation for arbitrary inputs. We divide the training dataset into 90% training and 10% validation. The model is trained with a batch size of 8. After hyperparameter tuning, the loss function we use in model training is the 90% relative L1 loss plus 10% relative L2 loss. The incorporation of L1-norm loss is more resistant to outliers. We use an Adam optimizer [49] with a learning rate of $10^{-3}$ and a weight decay of $10^{-5}$. We trained for 100 epochs, which takes approximately 40 minutes per epoch using a single NVIDIA Tesla V100 GPU with 24GB memory usage. A 70% of loss decrease is achieved in the first 10 epochs. Once the U-NO model is trained, the model parameters require GPU usage of 3.8 GB and the time for an evaluation on a new source and velocity model takes only 0.02 sec with GPU usage of 0.9 GB.

III. RESULTS

*A. The number of simulations needed for training*

Once completely trained, the U-NO model can be evaluated on a new input with very little computational cost (0.02 sec compared to the FDM runtime of 1.23 sec). The number of training simulations is the main factor in the computational cost. In the training process, we split the entire training dataset to 90% for training and 10% for validation. We test the performance of the model on the velocity models out of the training data set. We can see that the U-NO model trained on a dataset of 5000 simulations can already predict the major phase arrivals, while increasing the dataset size from 5000 simulations to 20000 provides better fit to the amplitudes (Figure 2). With a training dataset of 20000 simulations, the validation and training loss are very close, indicating there is no overfitting of the training data.

*B. Generalizability to arbitrary velocity structure or discretization*

The U-NO model is trained on random velocity models generated with the von Karman correlation function, which can best mimic the Earth's heterogenous velocity distribution [45], [50]. We show by example that the U-NO model, although trained on random velocity models with some certain parameters, is applicable to arbitrary velocity models. These outcomes are in fact expected from theoretical grounds because most physical functions can be approximated to arbitrary accuracy by random fields.

Our first example is with velocity models from a von Karman-type random distribution, but with a different covariance function than the one used for the training data. We increase the roughness of the velocity structure by a factor of four by decreasing the correlation length of $V_S$ and $V_P/V_S$ to only one-fourth that of the training data. As shown in Figure 3, the wavefield snapshot has more coda than with the smoother models because of the scattering from increased heterogeneity. However, the coda waves are well modeled by U-NO when compared to the ground truth simulation by FDM.

The velocity models used in the training data do not have coherent structures with discontinuities as in the real Earth, but wavefields for such models can still be simulated with our method. As mentioned before, this is because discontinuous functions can be approximated to arbitrary accuracy by random fields. Figure 4 shows a simple model with a dipping 'slab' embedded in a homogeneous background. The slab has 20% higher $V_S$ and 5% lower $V_P/V_S$. The wavefield snapshots show that the reflections from the high velocity anomaly are clearly predicted by U-NO. A more complex example is shown in Figure 5, where a random subpanel of the Marmousi model, a 2D velocity model with complex vertical and horizontal structures used in exploration studies [51], is used. The reflected and refracted waves are very complicated due to the presence of folding and faulting, but the U-NO predictions still closely approximate the numerical solutions (Figure 5).

One of the most important advantages of a neural operator compared with a neural network is its mesh-free nature, since it intrinsically learns the mapping between function spaces. A model trained on a particular mesh can be evaluated on any other mesh, even at finer spacing. The Fourier layers may learn from and evaluate functions on any discretization because parameters are directly learned in Fourier space and resolving the functions in physical space is simply projecting on the basis [38]. The example in Figure 6 shows the U-NO trained on a grid of 64*64 nodes applied to an input velocity model with 160*160 nodes. Here both the input velocity model and the output wavefield can be seen at a much higher resolution, yet U-NO provides comparable prediction with the FDM solver. Note that if the resolution is increased by a factor of 2, a grid-based numerical solver like FDM takes about 6 times greater computational time; however, the evaluation using U-NO takes only about 2.5 times longer, providing additional computational efficiency.

We evaluate the overall generalization performance of the trained U-NO by performing a thousand random realizations on each of these cases. The distribution of the relative L2-norm misfit and cross-correlation coefficient are plotted in Figure 7. In the case of the Marmousi model, the extended tail of the histogram is attributed to the model's imbalanced complexity. In general, however, we see a very high cross-correlation coefficient (>0.95) between U-NO prediction and ground truth, confirming its robust generalizability.

*C. Application to full-waveform inversion*

One of the most important applications of wavefield simulations is in full-waveform inversion (FWI), which uses the full recorded waveform to image the Earth's interior. The adjoint-state method is the traditional approach for computing the gradients of an objective function with respect to parameters of interest [4], [5]. Neural operators are differentiable by design, which enables gradient computation with reverse-mode automatic differentiation. It has been shown that automatic differentiation and the adjoint approach are mathematically equivalent [28]. Hence, the trained U-NO model allows for convenient FWI and the associated speed and accuracy should depend only on the forward modeling part.

We demonstrate the inversion performance using the velocity structure of random subpanels in the Marmousi model [51]. The synthetic waveform data are simulated with FDM [47] using 14 sources distributed in a ring shape. In Figure 8, we use the true source location, receivers on all 64*64 grids and noise-free waveform data; the goal here is not to demonstrate resolution, but rather the computational accuracy of the method. We then invert for $V_P$ and $V_S$ simultaneously by starting with homogeneous initial $V_P$ and $V_S$ models and forward propagating the wavefield with the U-NO for each source. The misfit is defined by the mean square error between the forward modeled and true wavefield. The gradient of the misfit with respect to $V_P$ and $V_S$ can be computed through automatic differentiation. $V_P$ and $V_S$ are then iteratively updated with gradient descent for 100 iterations using the Adam optimizer [49] with a learning rate of 0.01. Each iteration takes only about 1.4 sec by taking advantage of U-NO forward computation. The results in Figure 8 show a relative L2-norm misfit between the true and inverted model of only 3%. This successful inversion, in turn, further validates the accuracy of forward modeling with U-NO.

Besides the fact that the inversion target velocity model is quite different from the smooth random fields in the training dataset, this experiment itself is difficult due to conventional problems in full-waveform inversion, such as cycle skipping (multiple local maxima in the least-squares misfit function). We also show that if we only use 64 receivers on the surface, the inversion results in the region with ray path coverage are still reasonably accurate (Figure 9). Inversion with a biased homogenous initial model is also capable of producing relatively accurate results (see Supplementary Materials).

IV. DISCUSSION AND CONCLUSION

We use the relative L2 loss between the FDM and U-NO predictions to evaluate the performance of the trained model for generalization. The relative L2 loss is defined as the L2-norm of the difference between the prediction and ground truth divided by the L2-norm of the ground truth. This ratio is used to evaluate the performance of the trained model. When using the same mesh discretization as the training data, the relative L2 loss is around 10–20 percent, but this number rises to 30–40 percent when the tests are performed on finer grids (Figure 7). These values are misleading, however, because the relative L2 loss imposes equal weights to the entire sparse matrix of waveforms that is dominated by small amplitudes close to 0. Alternatively, the cross-correlation coefficient is a quantity that is more sensitive to the seismic phases with amplitudes larger than background noise. A cross-correlation coefficient larger than 0.95 suggests the coherence of the U-NO prediction is excellent, even for the scenarios with large relative L2 loss (Figure 7). In addition, the FWI results confirm that the large L2 loss is not so important since even challenging models can still be properly recovered (Figure 8).

Besides the more than an order of magnitude higher speed, the most important advantage of the neural operator-based full waveform modeling is its generalizability to arbitrary velocity models or discretization. This is because the neural operator learns a general solution operator to the wave equation instead of a specific instance of input velocity models. Once the neural operator is trained, it can be used by the entire seismology community for any region of a similar size without the need for retraining. Since the full waveform modeling with a neural operator has easily accessible gradients for convenient FWI, we anticipate that this approach will eventually make FWI as affordable as travel time tomography.

One of the main limitations of the method is the domain extent. For a trained neural operator, the function is defined on a fixed domain extent (e.g. it could be a unit cube). We can evaluate at a different grid size but cannot change the extent. We are now working on an extension of the work, where we recursively predict the wavefield. Through this way, a trained neural operator is essentially taking the first few time steps as input and then output the next few time steps, and there will be no need for retraining.

The scalability of evaluation using a trained neural operator with respect to the grid size and the number of time steps is a little different from conventional FDM. Assuming the original dimension is $(Nx_1, Ny_1, Nt_1)$, where $Nx_1, Ny_1$ are the number of grids in the x and y domain, respectively, and $Nt_1$ is the number of time steps. If the new dimension is $(Nx_2, Ny_2, Nt_2)$, the memory becomes $\frac{Nx_2 \cdot Ny_2 \cdot Nt_2}{Nx_1 \cdot Ny_1 \cdot Nt_1}$ times the original memory, which is consistent with the FDM. In the example presented in this paper, evaluation using U-NO takes 3 times the GPU memory of the FDM approach, and this scaling should be consistent with increasing grid points. In terms of computational cost, the majority of it for UNO is on the Fourier transform and its inverse. The computational cost of fast Fourier transform with dimension $(Nx_1, Ny_1, Nt_1)$ is proportional to $Nx_1 \cdot Ny_1 \cdot Nt_1 \cdot \log(Nx_1 \cdot Ny_1 \cdot Nt_1)$, and therefore, the new computational time becomes $\frac{Nx_2 \cdot Ny_2 \cdot Nt_2 \cdot \log(Nx_2 \cdot Ny_2 \cdot Nt_2)}{Nx_1 \cdot Ny_1 \cdot Nt_1 \cdot \log(Nx_1 \cdot Ny_1 \cdot Nt_1)}$ times the original computational time. This scaling is slightly higher than that of FDM, however, considering the 60 times acceleration in the example presented in this paper, UNO evaluation on an increased dimension of 1024*1024*1024 should still have ~40 times the acceleration.

The most compute- and memory-intensive part of the UNO method is the one-time training process. The cost of training for the 2D case is tractable on a single GPU. For the extension from 2D to 3D modeling, the computation and memory will increase due to the larger dataset and the larger number of parameters to learn. Therefore, the next step is to enhance data compression and parallelization to accelerate the training process and reduce the storage. Since this is a learning-based approach, the model performance can be improved by fine-tuning the model parameters and increasing the size of the training dataset. More importantly, any future advancements made in neural operator model architectures will be able to be directly incorporated into the system as they occur. For example, the improvement from linear layers of FNO to U-NO enables faster training convergence. As a result, we should only take current performance metrics as a starting point.

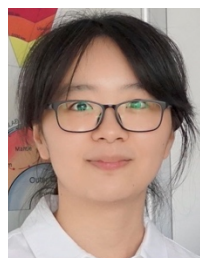

**Yan Yang** is a Ph.D. student with the Seismological Laboratory, California Institute of Technology, Pasadena, CA, USA. Her research interests focus on seismic imaging and monitoring of subsurface.

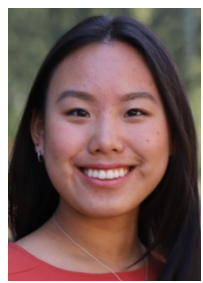

**Angela Gao** is a Ph.D. student with the Computing and Mathematical Sciences Department of the California Institute of Technology, Pasadena, CA, USA. She is interested in inverse problems, computational photography, generative models, and deep learning, with applications in scientific imaging problems.

**Kamyar Azizzadenesheli** is a Senior Research Scientist at Nvidia. Prior to his role at Nvidia, he was an assistant professor at Purdue University, department of computer science. Prior to his faculty position, he was at the California Institute of Technology as a Postdoctoral Scholar in the Department of Computing and Mathematical Sciences.

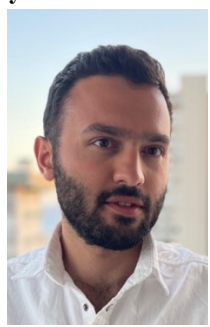

**Robert W. Clayton** is a Professor of geophysics at California Institute of Technology, Pasadena, CA, USA, where he works in the areas of seismic wave propagation, earth structure, and tectonics. He has applied imaging methods to the Los Angeles region and to subduction zones around the world.

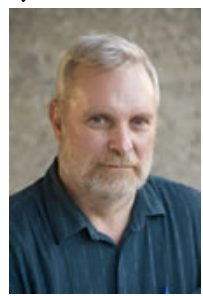

**Zachary E. Ross** is an Assistant Professor of geophysics with the California Institute of Technology, Pasadena, CA, USA, where he uses machine learning and signal processing techniques to better understand earthquakes and fault zones. He is interested in the dynamics of seismicity, earthquake source properties, and fault zone imaging.

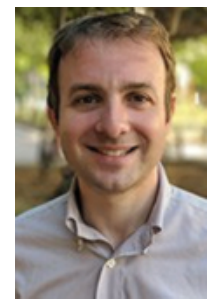

# Figures

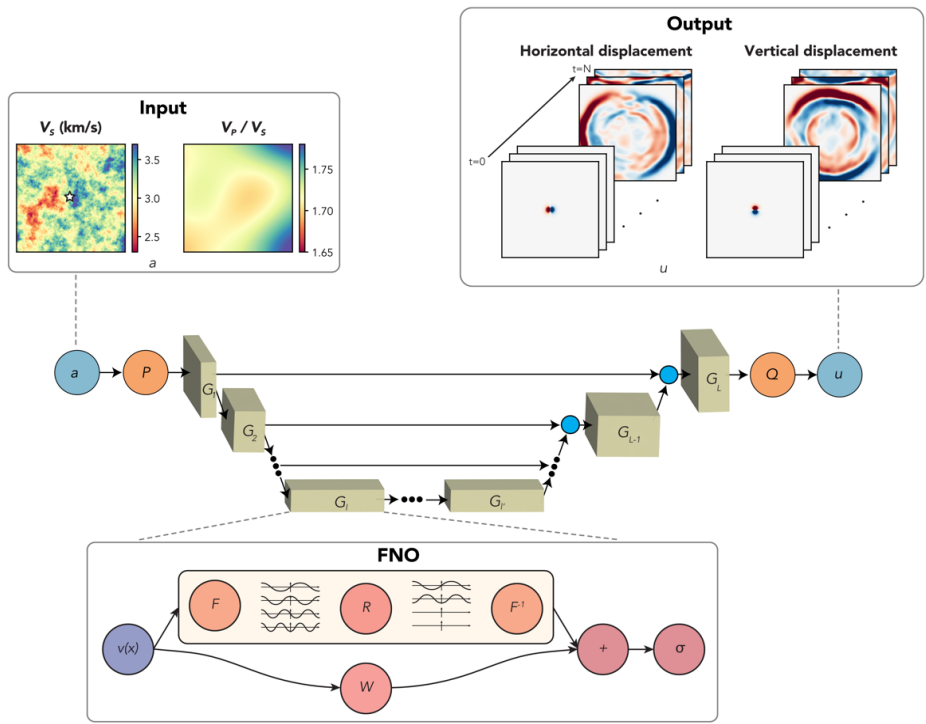

**Figure 1. Overview of our method for solving the elastic wave equation with neural operators.** The inputs, *a*, to the U-NO model are the P- and S-wave velocity ($V_P$, $V_S$) model and the source location (indicated by the white star). $V_P$ is calculated from $V_S$ and $V_P/V_S$, examples of which are shown in the upper left panel. The outputs, *u*, are the horizontal and vertical displacements at each time step, examples of which are shown in the upper right panel. In the middle panel showing the U-NO architecture, orange circles *P* and *Q* denote point-wise operators, rectangles *G* denote general operators, and smaller blue circles denote concatenations in function space. The lower panel shows the architecture of each FNO layer, where *v* is the input of the layer, *F* and $F^{-1}$ are Fourier transform and its inverse, respectively, R and W are a linear transform, and $\sigma$ is the nonlinear activation function. The expansion (or contraction) factors in equation (5) are set as: $c_{1,2}^s = \frac{3}{4}, c_{3,4}^s = \frac{1}{2}, c_{5,6}^s = 2, c_{7,8}^s = \frac{4}{3}, c_1^t = \frac{3}{4}, c_2^t = \frac{2}{3}, c_{3,4}^t = \frac{1}{2}, c_{5,6}^t = 2, c_7^t = \frac{3}{2}, c_8^t = \frac{4}{3}, c_{1,2,3,4}^c = 2, c_{5,6,7,8}^c = \frac{1}{2}$.

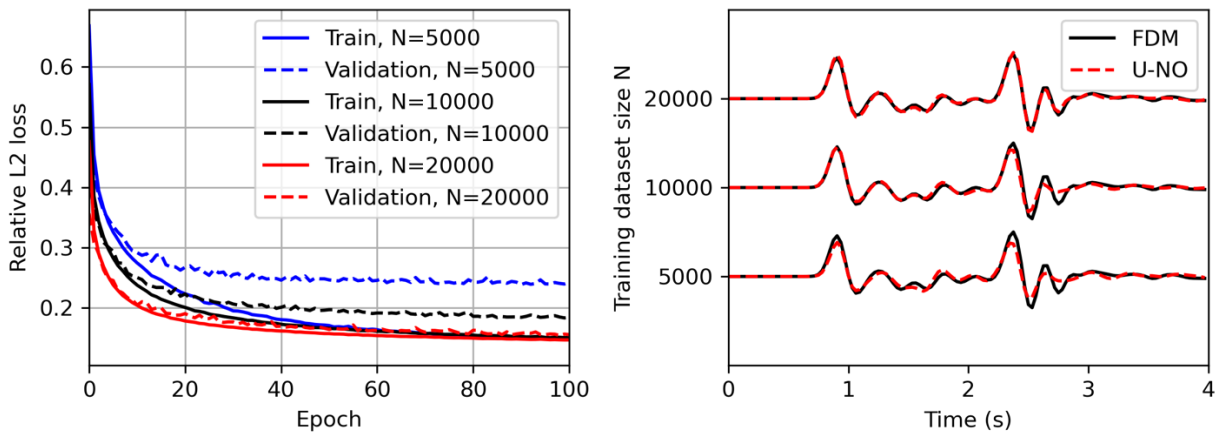

**Figure 2 Model performance as a function of the number of training samples.** Left panel: relative L2 loss curves for the training and validation data. Right panel: Example of simulated waveform comparison between FDM (black solid) and U-NO (red dashed). N means the number of simulations in the training dataset, including 90% for training and 10% for validation.

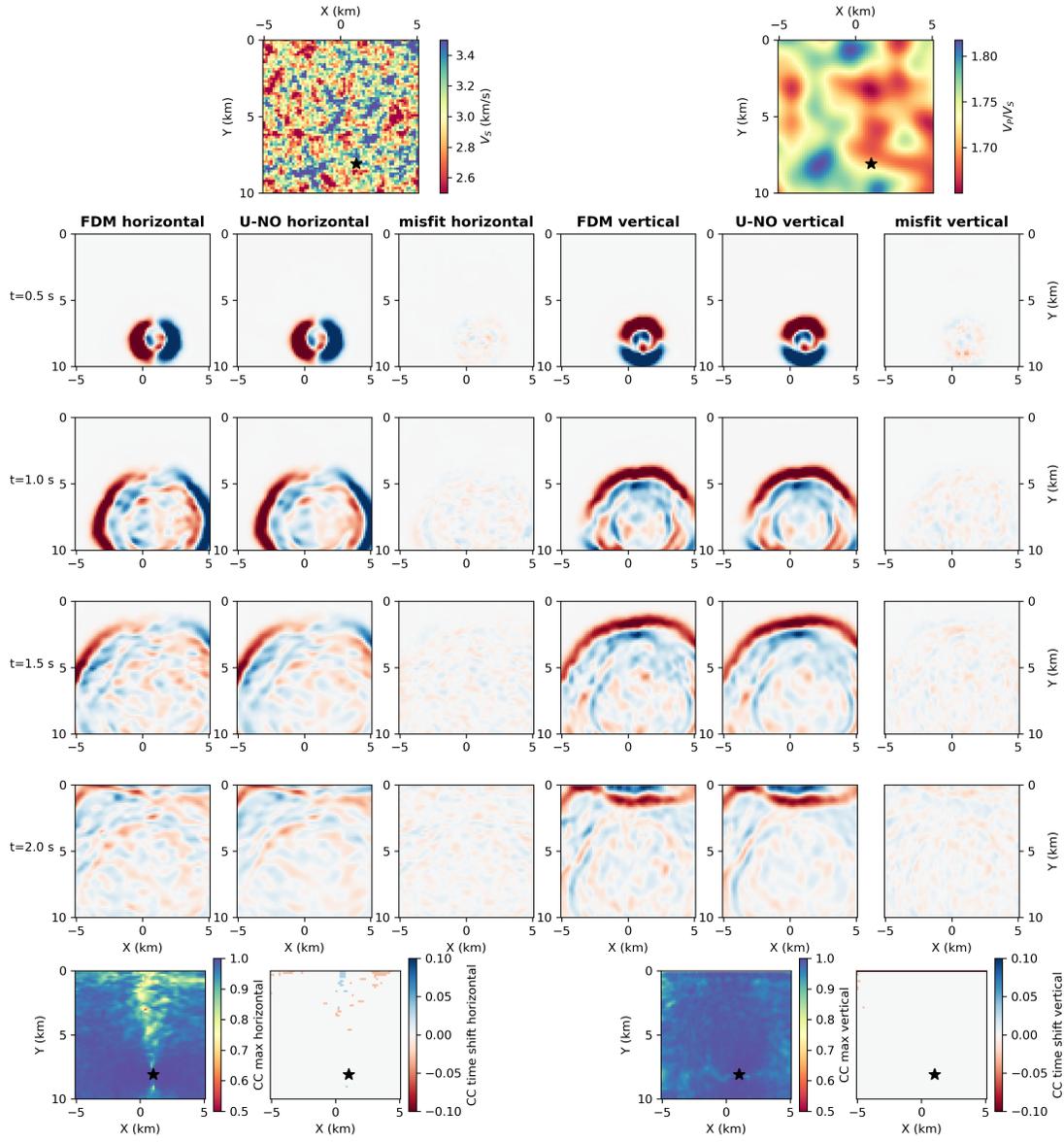

**Figure 3 Model generalization experiments 1**: random fields of Vs and Vp/Vs model with 4-times roughness of the training data. The top row shows the $V_S$ and $V_P/V_S$. From the second row to the fifth row, the wavefield snapshots at 0.5 s-2.0 s are shown. From left to right, the first three columns show the horizontal displacement of the FD simulation, U-NO prediction, and their misfit in the same color scale. The latter three columns show the vertical component. The horizontal and vertical displacement waveforms at each grid are cross-correlated between FD and U-NO, with the maximum cross-correlation value and its associated time shift shown in the bottom row. For this case, the relative L2 loss of the U-NO simulation is 0.182.

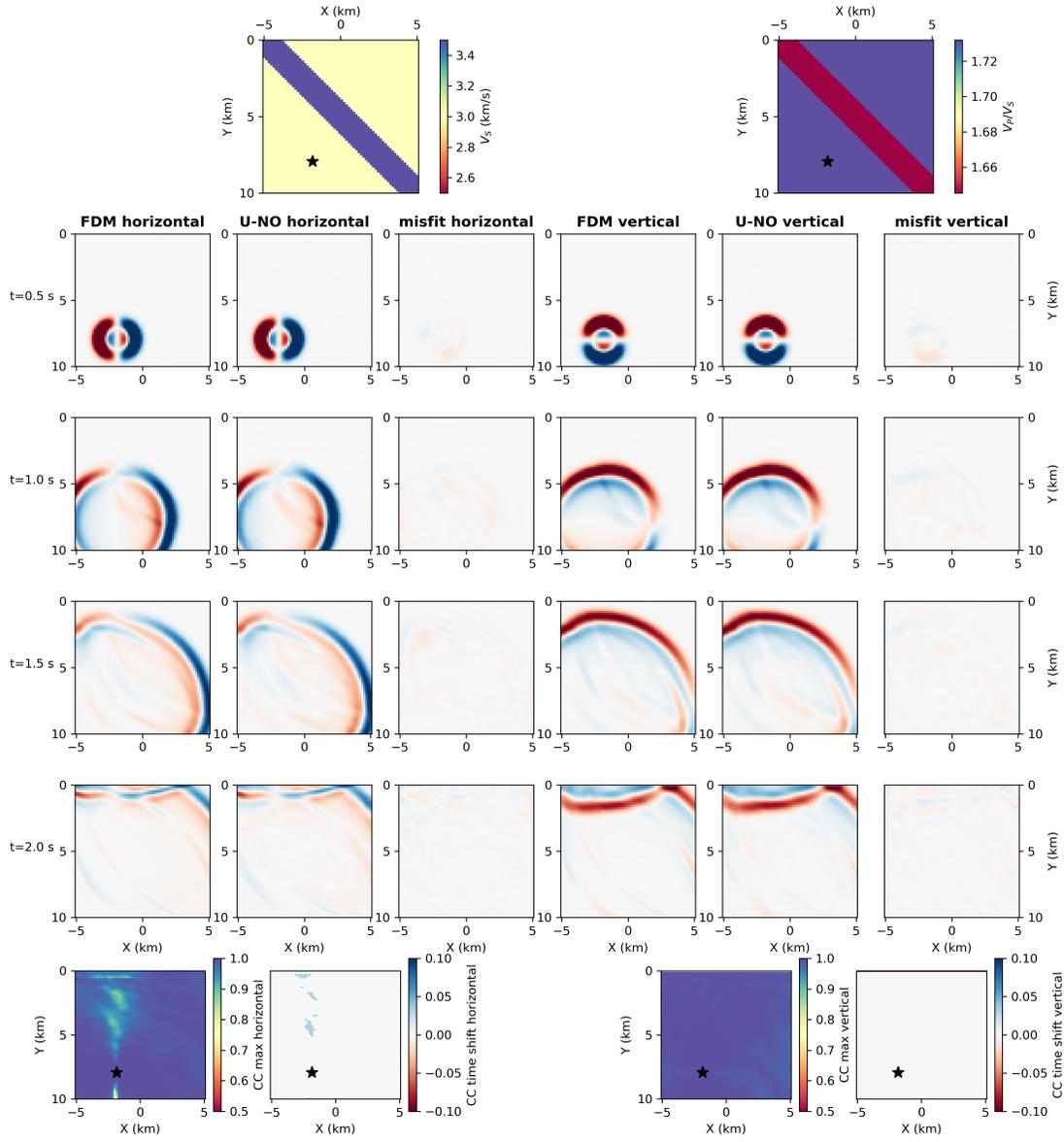

**Figure 4 Model generalization experiments 2:** Similar as Figure 3, but for a homogeneous background model embedded with a 'slab' with 20% higher Vs and 5% lower Vp/Vs. Relative L2 loss of the U-NO simulation is 0.090.

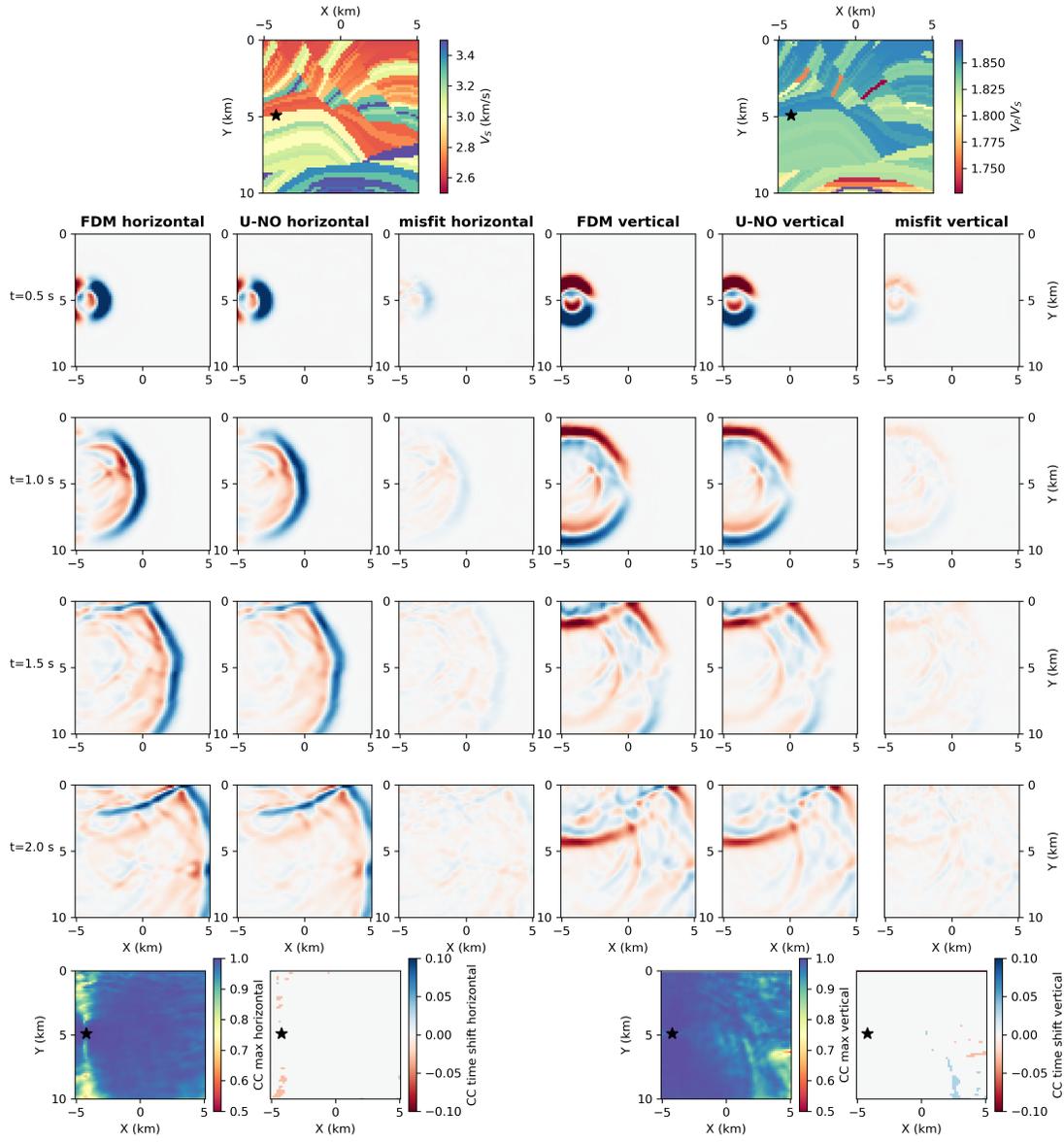

**Figure 5 Model generalization experiments 3:** Similar as Figure 3, but for a random subpanel from the Marmousi model. The velocity perturbation range is normalized to 30% of the average velocity. Relative L2 loss of the U-NO simulation is 0.225.

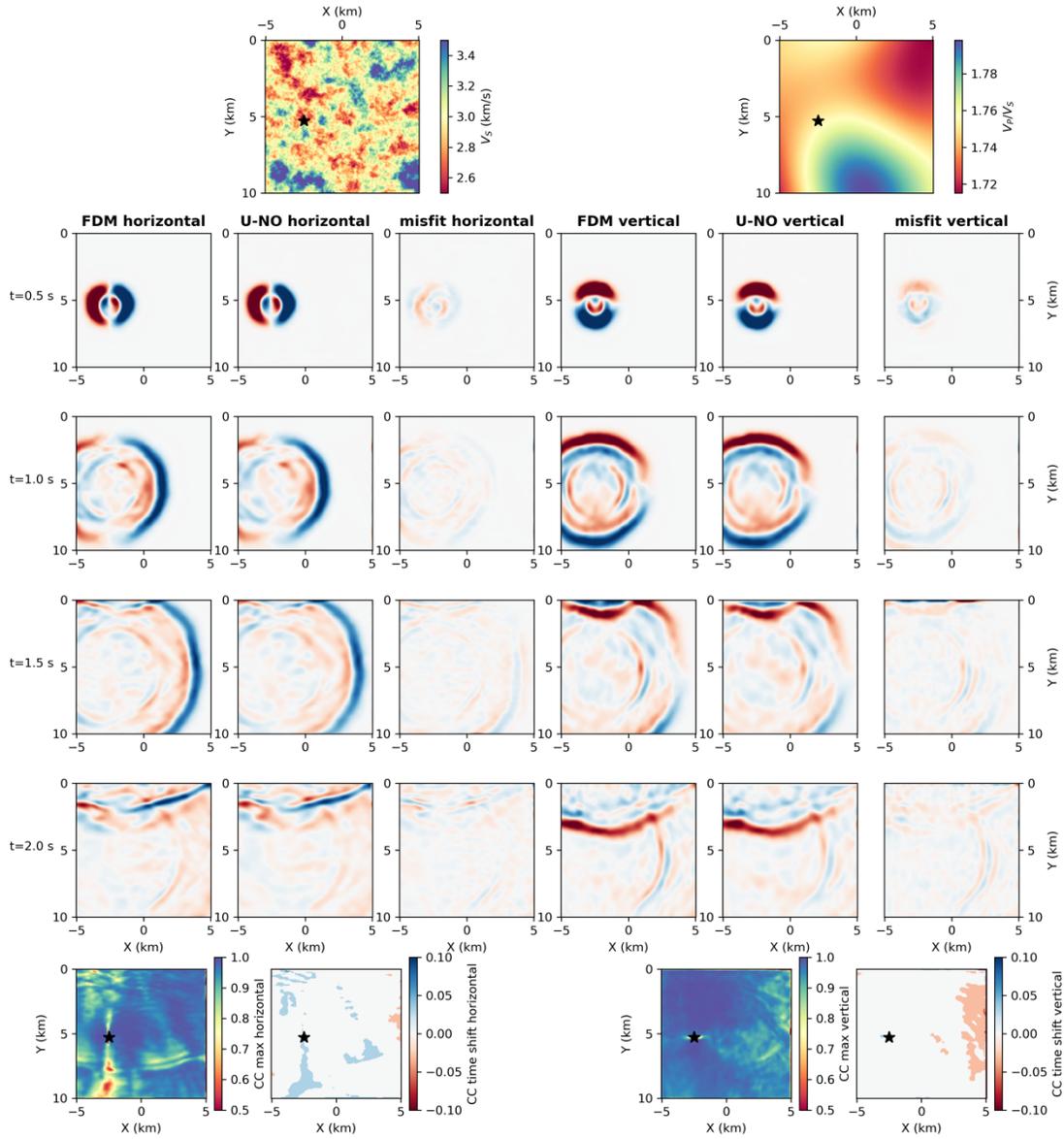

**Figure 6 Model generalization experiments 4:** Similar as Figure 3, but Vp and Vs model mesh discretization is increased from 64*64 to 160*160. Relative L2 loss of the U-NO simulation is 0.385.

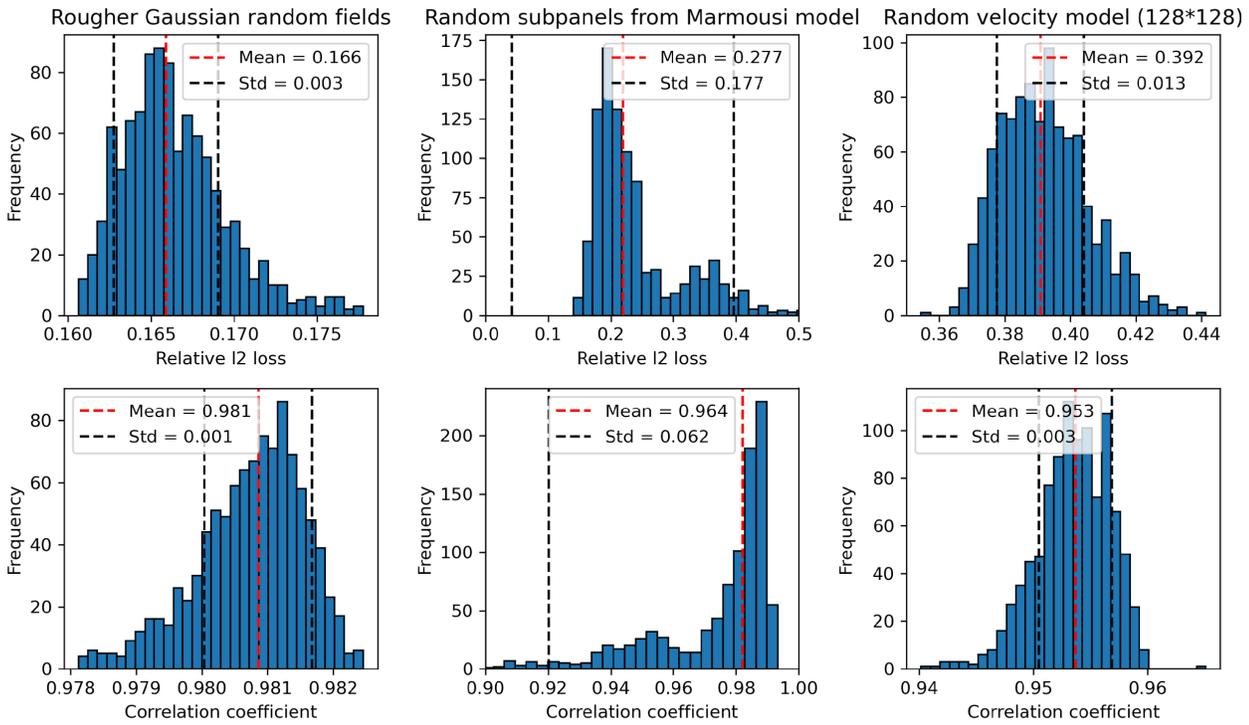

**Figure 7 Output evaluation.** Distribution of Relative L2 loss (top) and correlation coefficient (bottom) between the U-NO predictions and ground truth. From left to right, the columns are corresponding to the experiments in Figure 3, 5, 6. The red and black dashed lines mark the mean and standard deviation of the histograms.

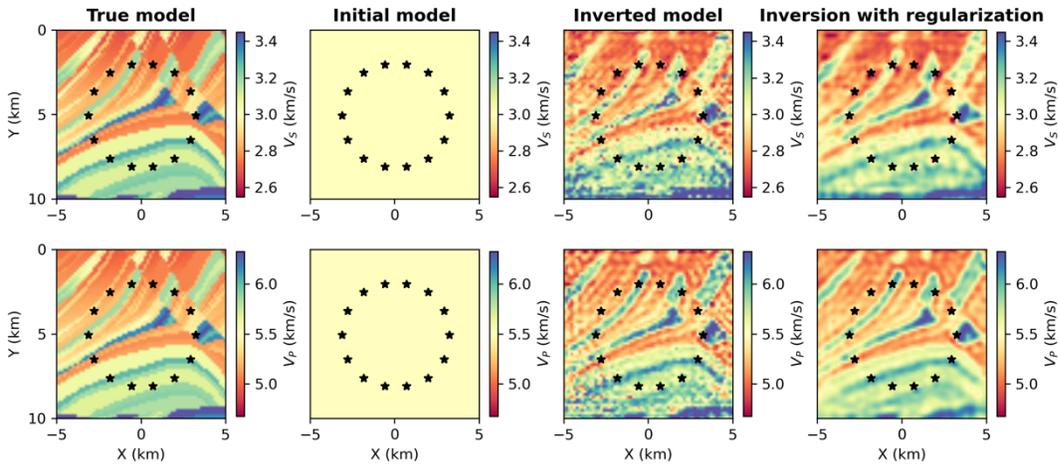

**Figure 8 Full waveform inversion.** The inversion for a random subpanel from the Marmousi model. 14 sources are placed in a ring shape (black stars) and receivers are placed at every node of the 64*64 grid. From left to right, the columns represent true velocity model, initial model for inversion, inverted model without regularization, and inverted model with $0^{th}$ and $1^{st}$ order Tikhonov regularization. The top and bottom rows are the models for Vs and Vp, respectively.

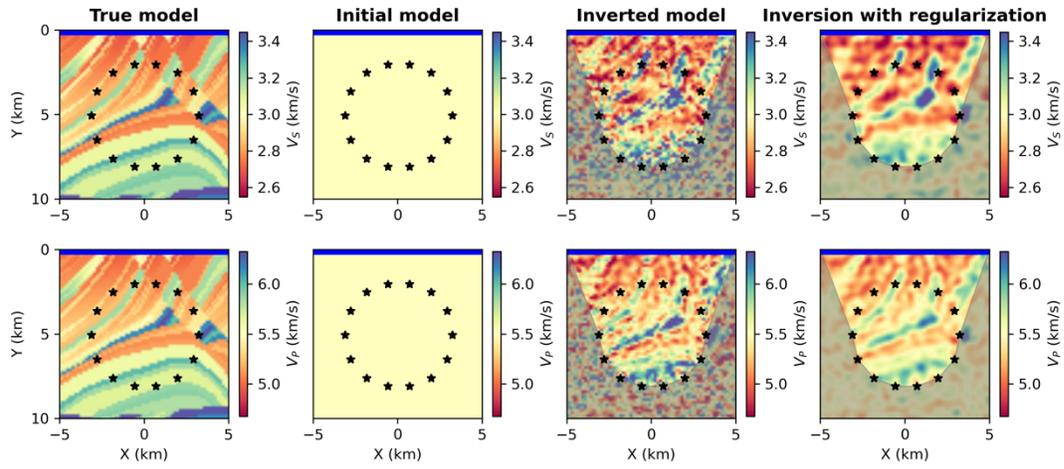

**Figure 9 Full waveform inversion.** Same as Figure 8, but the receivers are only placed on the 64 grids on the surface (blue line on the top of each subpanel). The gray shaded areas mask the areas without ray path coverage.